\documentclass[preprint,12pt]{elsarticle}

\usepackage{amssymb}
\usepackage{amsmath,amsfonts}
\usepackage{algorithmic}
\usepackage{algorithm}
\usepackage[group-separator={,}, group-minimum-digits={3}]{siunitx}
\usepackage{tabularx}
\usepackage{multirow}
\usepackage{booktabs}
\usepackage{makecell}
\usepackage{subfig}
\usepackage{setspace}

\journal{Pattern Recognition}

\begin{document}

\begin{frontmatter}



\title{Detecting Political Opinions in Tweets through Bipartite Graph Analysis: A Skip Aggregation Graph Convolution Approach}

\author[label1]{Xingyu Peng}
\author[label2]{Zhenkun Zhou\corref{cor1}}
\ead{zhenkun@cueb.edu.cn}
\cortext[cor1]{Corresponding author}
\author[label1]{Chong Zhang}
\author[label1]{Ke Xu}
\affiliation[label1]{
    organization={State Key Lab of Software Development Environment, Beihang University},
    city={Beijing},
    postcode={100191},
    country={China}
}

\affiliation[label2]{
    organization={School of Statistics, Capital University of Economics and Business},
    city={Beijing},
    postcode={100070},
    country={China}
}

\begin{abstract}
Public opinion is a crucial factor in shaping political decision-making. Nowadays, social media has become an essential platform for individuals to engage in political discussions and express their political views, presenting researchers with an invaluable resource for analyzing public opinion. In this paper, we focus on the 2020 US presidential election and create a large-scale dataset from Twitter. To detect political opinions in tweets, we build a user-tweet bipartite graph based on users' posting and retweeting behaviors and convert the task into a Graph Neural Network (GNN)-based node classification problem. Then, we introduce a novel skip aggregation mechanism that makes tweet nodes aggregate information from second-order neighbors, which are also tweet nodes due to the graph's bipartite nature, effectively leveraging user behavioral information. The experimental results show that our proposed model significantly outperforms several competitive baselines. Further analyses demonstrate the significance of user behavioral information and the effectiveness of skip aggregation.
\end{abstract}



\begin{keyword}
political opinion mining \sep graph neural networks \sep social media



\end{keyword}

\end{frontmatter}


\section{Introduction}
In today's digital age, social media has emerged as an indispensable tool for the general public to engage in discussions on political issues, such as elections, taxes, education, and regulations. As was seen in the 2016 US presidential election \cite{bovet2018validation} and the 2019 Argentina presidential election \cite{zhou2021polls}, social media provides a platform for candidates to engage with potential voters, share their policies, and build their brand during the pre-election phase. Similarly, voters can use social media to learn more about the candidates, their policies, and their positions on important issues, as well as express their opinions on the election. With the occurrence of such political events, significant numbers of reviews and responses surface online, making it possible to capture public opinion and social trends. However, manually analyzing such a massive amount of textual data is extremely time-consuming and costly, so it is imperative to automate the analysis procedure. 

Early studies have primarily concentrated on people's opinions about hotels, restaurants, electronics, and other consumer products. In contrast, understanding opinions in a political context \cite{manickam2019ideotrace, xiao2020timme, fagni2022fine, maynard2011automatic, li2019encoding, li2021using, feng2021knowledge, zhang2022kcd, iyyer2014political, chen2017opinion, xiao2022detecting} has received less attention, and is generally more difficult due to nuanced language, ambiguous sentiment, and complex context. In this study, our focus is on the 2020 US presidential election, and we aim to ascertain the political polarities of election-related tweets that were posted during the pre-election period. Compared to news articles \cite{li2019encoding, li2021using, feng2021knowledge, zhang2022kcd}, debate transcripts \cite{iyyer2014political, chen2017opinion}, or tweets from official sources like legislators, news agencies, and politicians \cite{xiao2022detecting}, which typically follow strict grammatical rules, analyzing tweets posted by ordinary users presents unique challenges, such as the brevity of the text, absence of contextual information, and frequent use of emoticons, abbreviations, and hashtags. 

\begin{figure}[t]
  \begin{minipage}[t]{0.5\linewidth}
    \centering
    \includegraphics[width=0.88\linewidth]{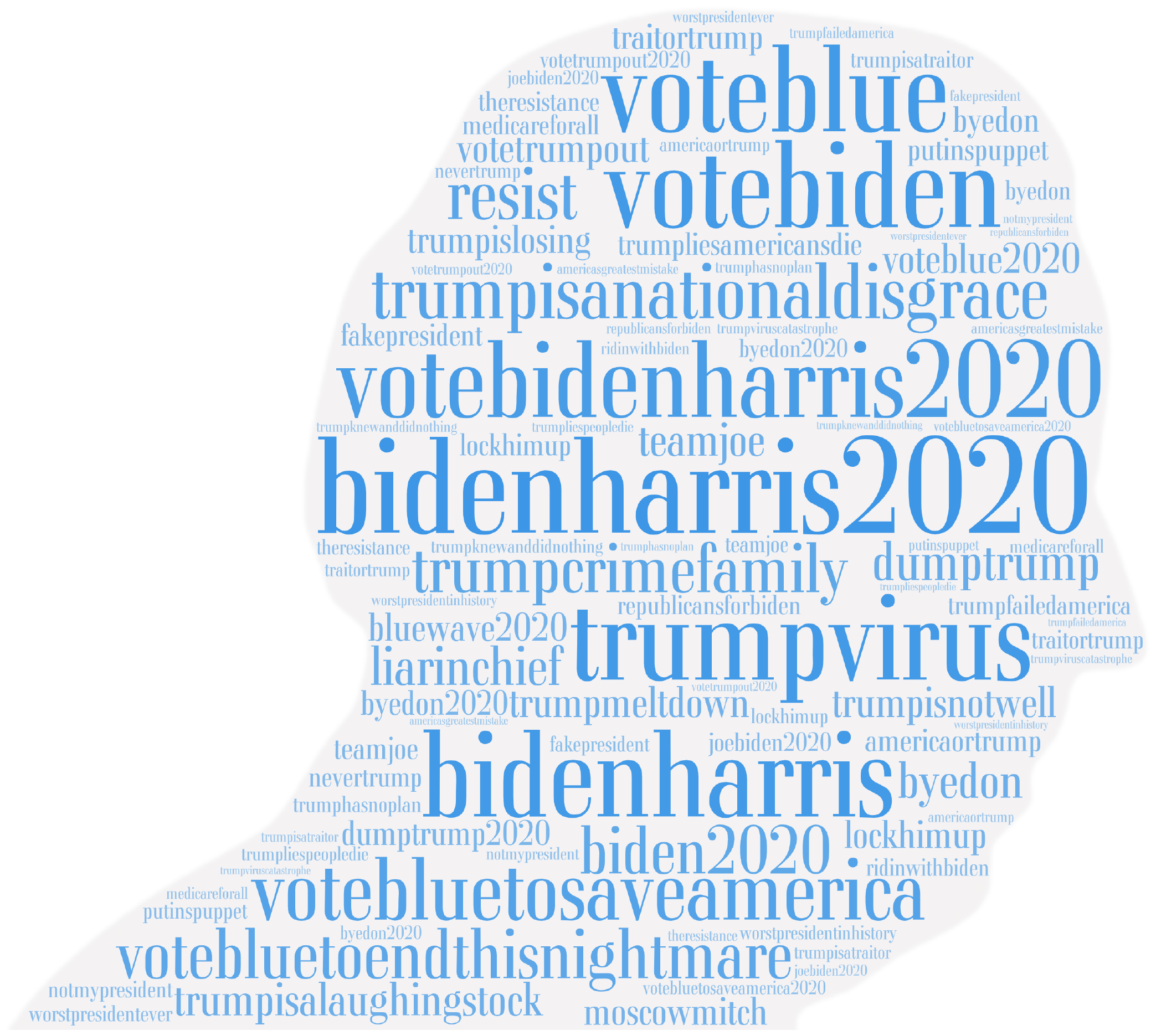}
  \end{minipage}%
  \begin{minipage}[t]{0.5\linewidth}
    \centering
    \includegraphics[width=0.80\linewidth]{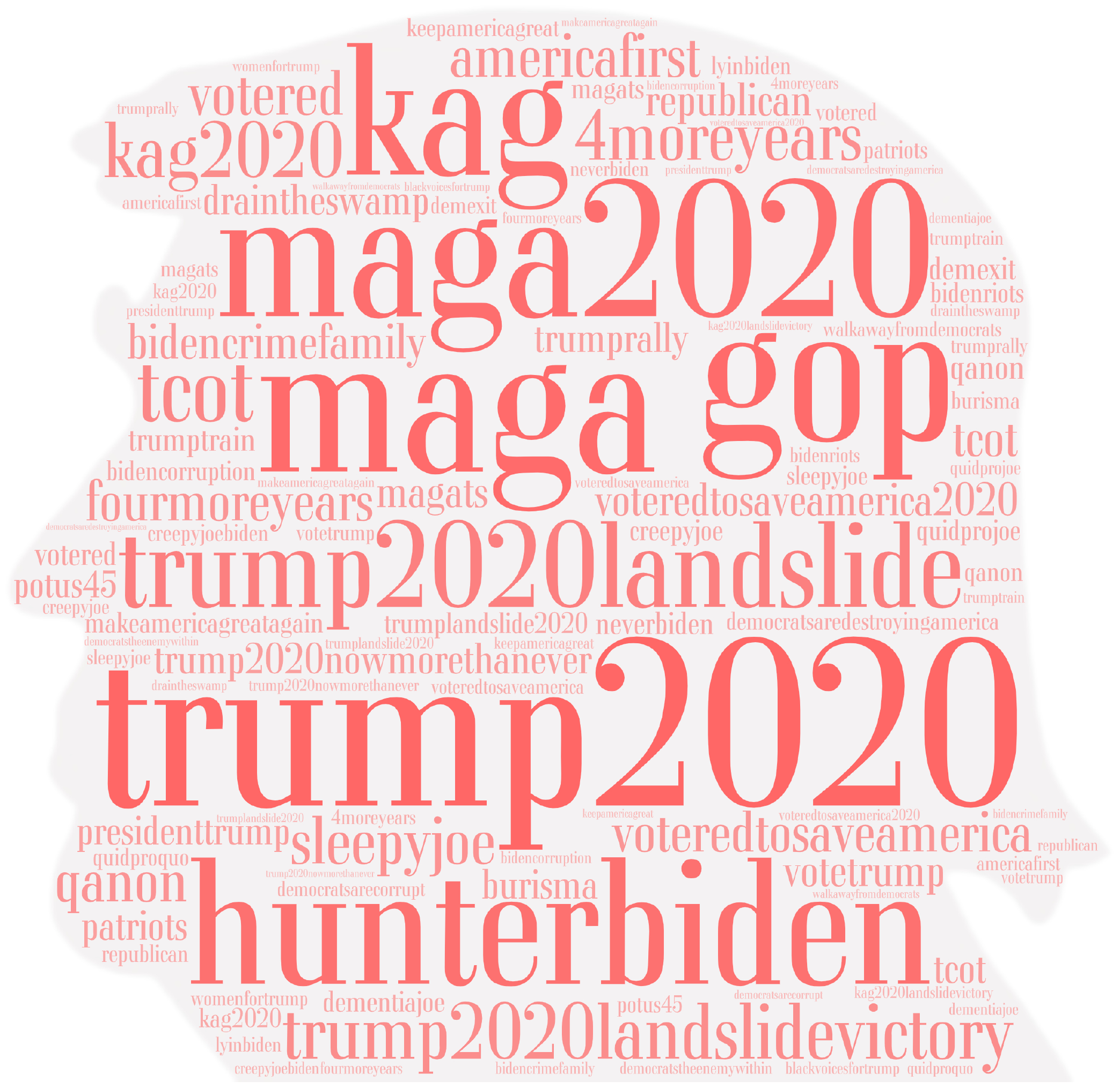}
  \end{minipage}
  \caption{Hashtag clouds. The words or multi-word phrases in the clouds are the most frequently occurring hashtags related to the 2020 US presidential election. The hashtags in blue express support for Biden or opposition to Trump (pro-Biden), while the hashtags in red express support for Trump or opposition to Biden (pro-Trump). The dimension of the hashtags is proportional to their frequency.
}
  \label{fig:hashtags}
\end{figure}

Additionally, labels for such social data are difficult to obtain. Because of the vast amount of data, manual annotation methods such as crowdsourcing and editorial review \cite{li2019encoding} are impractical. Some studies \cite{xiao2022detecting, iyyer2014political, chen2017opinion} annotate texts based on authors' or speakers' ideological positions, whereas the political ideologies of personal Twitter accounts are generally unknown. Actually, hashtags are frequently used by Twitter users to provide context and convey their sentiments or opinions on particular topics or issues. For example, a user might include a hashtag like \textit{\#ActOnClimate} to advocate for action on climate change, or \textit{\#BlackLivesMatter} to express support for the movement to end police brutality and racism against Black people. In general, hashtags tend to align with the overall opinion of the tweet they are included in. Therefore, we manually select a set of hashtags expressing support or opposition to one of the candidates and categorize them as either pro-Biden or pro-Trump. In cases where a hashtag opposes one candidate, we consider it as supporting the other candidate (e.g., treating \textit{\#traitortrump} as pro-Biden). We then annotate the tweets according to the polarities of the hashtags within them. The hashtags used for tweet annotation are presented in Figure~\ref{fig:hashtags}.

So far, we have acquired a sizable set of labeled tweets, which can help the model overcome the challenge of informal language and comprehend the rich context of the election. Instead of relying solely on text analysis to infer the opinions expressed in these tweets, we propose leveraging the abundance of behavioral data available on social media platforms. Specifically, we first build a user-tweet bipartite graph, where users and tweets are represented as nodes, and edges denote users' posting and retweeting behaviors towards tweets. Then, we conduct the task of political opinion mining as a GNN-based node classification problem. We initialize tweet node features using a pre-trained language model (e.g., BERT \cite{devlin2018bert}). However, the features of user nodes cannot be initialized in the same way due to the lack of textual information, making it difficult to assign them appropriate initial features aligned with tweet nodes. To address this issue, we introduce a novel skip aggregation mechanism that leverages the bipartite nature of the graph. At each iteration, every tweet node aggregates information from its second-order neighbors, which comprise other tweets posted or retweeted by its author or retweeters and tend to express similar opinions. User nodes only play a bridging role during neighborhood sampling and do not require any characterization. In addition, we incorporate edge type information into the model to make more effective use of behavioral data, which leads to further improvements in performance.

We compare our model with several existing competitive text classification and node classification models. The experimental results show that our model significantly outperforms these models, achieving the best performance. Further analyses demonstrate the significance of incorporating user behavioral information in political opinion mining and the effectiveness of skip aggregation in leveraging such information. Moreover, we provide evidence of our model's robustness in handling short texts through performance analysis and showcase the high-quality tweet representations learned by our model through embedding visualization. Our contributions can be summarized as follows:
\begin{itemize}
    \item We collect a large amount of data from Twitter related to the 2020 US presidential election, including tweets and associated metadata such as authors and retweeters, and automatically annotate the tweets using manually labeled hashtags.
    \item We propose a GNN-based framework to detect political opinions in tweets. The framework adopts a novel skip aggregation mechanism to effectively learn meaningful tweet representations from a bipartite graph consisting of user-tweet interactions.
    \item We conduct extensive experiments and analyses to demonstrate the significance of user behavioral information and the effectiveness of skip aggregation, both of which contribute to significantly better performance of our model over baselines. 
\end{itemize}

\section{Related Work}
\subsection{Political Opinion Mining}
Political opinion mining is a Natural Language Processing (NLP) task that involves analyzing political texts to identify the author's stance or opinion on a certain political event or issue. This is an important area of research that can provide insights into public opinion and help inform policy decisions. 
Researchers have conducted numerous studies in this area, including but not limited to determining the political leanings of tweets \cite{maynard2011automatic, iyyer2014political, chen2017opinion, xiao2022detecting}, inferring the political alignments of Twitter users \cite{manickam2019ideotrace, xiao2020timme, fagni2022fine}, detecting the amount of polarisation in the electorate \cite{conover2011political}, and predicting voting intentions or election results \cite{lampos2013user, zhou2021polls}. Maynard and Funk \cite{maynard2011automatic} employed advanced NLP techniques and incorporated extra-linguistic contextual information to extract more meaningful and higher quality opinions from a collection of pre-election tweets. Rather than relying on traditional bag-of-words methods that overlook the syntactic structure, Iyyer \textit{et al.} \cite{iyyer2014political} applied Recursive Neural Network (RNN) to accurately identify the political position evinced by a sentence. Chen \textit{et al.} \cite{chen2017opinion} suggested first building an opinion-aware knowledge graph by integrating extracted opinions and targeted entities into an existing structured knowledge base, and then performing ideology inference by propagating information across the graph. Xiao \textit{et al.} \cite{xiao2022detecting} proposed to quantify the political polarity of tweets by explicitly assigning polarity scores to the entities and hashtags within them. Manickam \textit{et al.} \cite{manickam2019ideotrace} introduced a framework for jointly estimating the ideology of social media users and news websites, and tracing changes in user ideology over time. Xiao \textit{et al.} \cite{xiao2020timme} presented a framework for ideology detection on Twitter, which utilizes a multi-relational encoder and a multi-task decoder to assess the significance of each relation. Fagni and Cresci \cite{fagni2022fine} introduced an unsupervised method for extracting fine-grained political leanings from social media posts. The method projects users into a low-dimensional ideology space, where they are clustered and their political leanings are automatically derived from the assigned cluster. Conover \textit{et al.} \cite{conover2011political} employed a two-phase approach to tackle the problem of predicting the political alignments of users. Lampos \textit{et al.} \cite{lampos2013user} analyzed tweets from the UK and Austria and demonstrated successful prediction of voting intention in over 300 polls across both countries. Zhou \textit{et al.} \cite{zhou2021polls} proposed an opinion tracking method that utilizes machine learning models and social network big data analysis to overcome the limitations of traditional polls and achieve accurate results in the 2019 Argentina elections.

In addition to analyzing social media data, researchers have also investigated methods for detecting political perspectives in news media. Li and Goldwasser \cite{li2019encoding} suggested that the political perspectives expressed in news articles can be inferred from the way the documents are disseminated and the characteristics of the users who endorse them. They then applied GCN to capture the social information embedded in the spread of news articles. Li and Goldwasser \cite{li2021using} introduced a framework for pre-training models that leverages the rich social and linguistic context readily available in data sources. Li and Goldwasser \cite{li2021mean} proposed an entity-centric framework that incorporates entity and relation representations learned from external knowledge sources and text corpus and utilizes word- and sentence-level attention mechanisms to measure the importance of different aspects of the article for prediction. Feng \textit{et al.} \cite{feng2021knowledge} framed the task as a graph classification problem. They constructed heterogeneous information networks to jointly model news content and external knowledge and adopted a relational GNN to learn graph representations. Zhang \textit{et al.} \cite{zhang2022kcd} proposed a novel approach that employs textual cues as paragraph-level labels and integrates multi-hop knowledge reasoning for inference.

\subsection{Graph Neural Networks}
GNNs have emerged as a powerful tool for modeling complex and structured data that can be represented as graphs. Graph Convolutional Network (GCN) \cite{kipf2016semi} utilizes the Laplacian matrix to represent the structure of the graph and capture spatial features between the nodes. GCN can be considered as an approximation of spectral domain convolution of graph signals. Graph Attention Network (GAT) \cite{velivckovic2017graph} combines the power of GCN with the attention mechanism, which allows nodes to selectively attend to their most informative neighbors. The graph convolutional operation used in models such as GraphSAGE \cite{hamilton2017inductive} and FastGCN \cite{chen2018fastgcn} can be viewed as a process of sampling and aggregating of neighborhood information, which is computationally efficient and scalable to large graphs. Graph Isomorphism Network (GIN) \cite{xu2018powerful} was developed to support more complex forms of aggregation and has been proven to be theoretically the most powerful GNN under the neighborhood aggregation scheme. GIN's learnable aggregation function allows it to capture more fine-grained information about the graph structure and learn more expressive representations of the nodes in the graph. Although GCN frameworks have shown great success in modeling single-relation graphs, they are limited to single-relation graphs and ignore the various types of relations that exist in realistic scenarios. Relational Graph Convolutional Network (R-GCN) \cite{schlichtkrull2018modeling} extends GCN to handle multi-relational graphs by introducing relation-specific transformations that consider the type and direction of each edge. R-GCN have achieved state-of-the-art results on a number of node classification and link prediction tasks for multi-relational graphs. The success of R-GCN highlights the importance of considering the different types of relations in complex graphs.

In recent years, GNNs have also been successfully applied in various NLP tasks including text classification, sentiment analysis, machine translation, and question answering. One of the key benefits of using GNNs in NLP is the ability to capture the structural information in text data, such as syntax and semantics, which can be represented as graphs. Yao \textit{et al.} \cite{yao2019graph} proposed TextGCN, which is a pioneering method that employs GCN for text classification. TextGCN involves creating a heterogeneous graph consist of both word nodes and document nodes and converting the task into a node classification problem. The results of TextGCN showed significant improvements over a variety of state-of-the-art methods and attracted much attention in the NLP community. Chen \textit{et al.} \cite{chen2019graphflow} introduced a GNN-based model for the conversational Machine Reading Comprehension (MRC) task. The model utilizes an attention mechanism to dynamically build a passage graph that incorporates both the question and the conversation history. Each passage word is represented as a node in the graph at each turn of the conversation. Xu \textit{et al.} \cite{xu2020document} proposed a novel hybrid graph model for document-level Neural Machine Translation (NMT) that considers both intra- and inter-sentential relations, which addresses the severe long-dependency issue that arises in document-level NMT by allowing the model to capture dependencies not only within sentences but also across sentence boundaries. Hu \textit{et al.} \cite{hu2021compare} developed an end-to-end graph neural model for fake news detection, which compares the news to the knowledge base through entities and incorporates topics to enrich the news representation. Peng \textit{et al.} \cite{ijcai2022p601} applied GNNs to the task of document-level Relation Extraction (RE). They extract a subgraph from the document graph by tracing reasoning paths between the target entity pair and use R-GCN to capture the relations.

\section{Methodology}
\subsection{Data Preparation, Annotation, and Preprocessing}
In this study, we focus on the 2020 United States presidential election held on November 3, 2020. We crawl tweets posted between October 1, 2020, and November 2, 2020, using the Twitter Streaming API, filtering by the following query: \textit{biden} $OR$ \textit{trump}, which corresponds to the two main candidates from the Democratic Party (Joe Biden) and the Republican Party (Donald Trump). In total, we gather 138.9 million tweets in English, along with the associated metadata for each tweet. 

We leverage the fact that users frequently include hashtags in their tweets, which tend to reflect the opinion expressed in the entire tweet, using hashtags to annotate the tweets we previously collected. In particular, we first identify the most commonly used hashtags and manually categorize those explicitly expressing support or opposition to one of the candidates as either pro-Biden or pro-Trump. For hashtags that oppose one candidate, we treat them as supporting the other candidate (e.g., \textit{\#traitortrump} is considered pro-Biden). Next, we expand the hashtag set by discovering new hashtags that are significantly related to the initial set, based on co-occurrence. As a result, the expanded hashtag set contains 245 labeled hashtags. We select a subset of these hashtags based on their frequency of occurrence and present them in Figure~\ref{fig:hashtags}. Then, we filter tweets that contain at least one hashtag from the hashtag set. In cases where a tweet contains multiple hashtags, we keep it only if all the hashtags support the same candidate. Finally, we annotate each tweet according to the polarity of the hashtags within it. 

To avoid duplicates, we remove retweets (tweets that start with ``RT @username'') and record the retweet information as metadata in the corresponding original tweets. Additionally, we remove the hashtags used for annotation from tweets to avoid target leakage, while retaining other hashtags as they may contain crucial information about the opinion. The corpus ultimately consists of \num{1123749} labeled tweets, which is sufficient to train an effective neural classifier.

\begin{figure}[t]
\centering
\includegraphics[width=\linewidth]{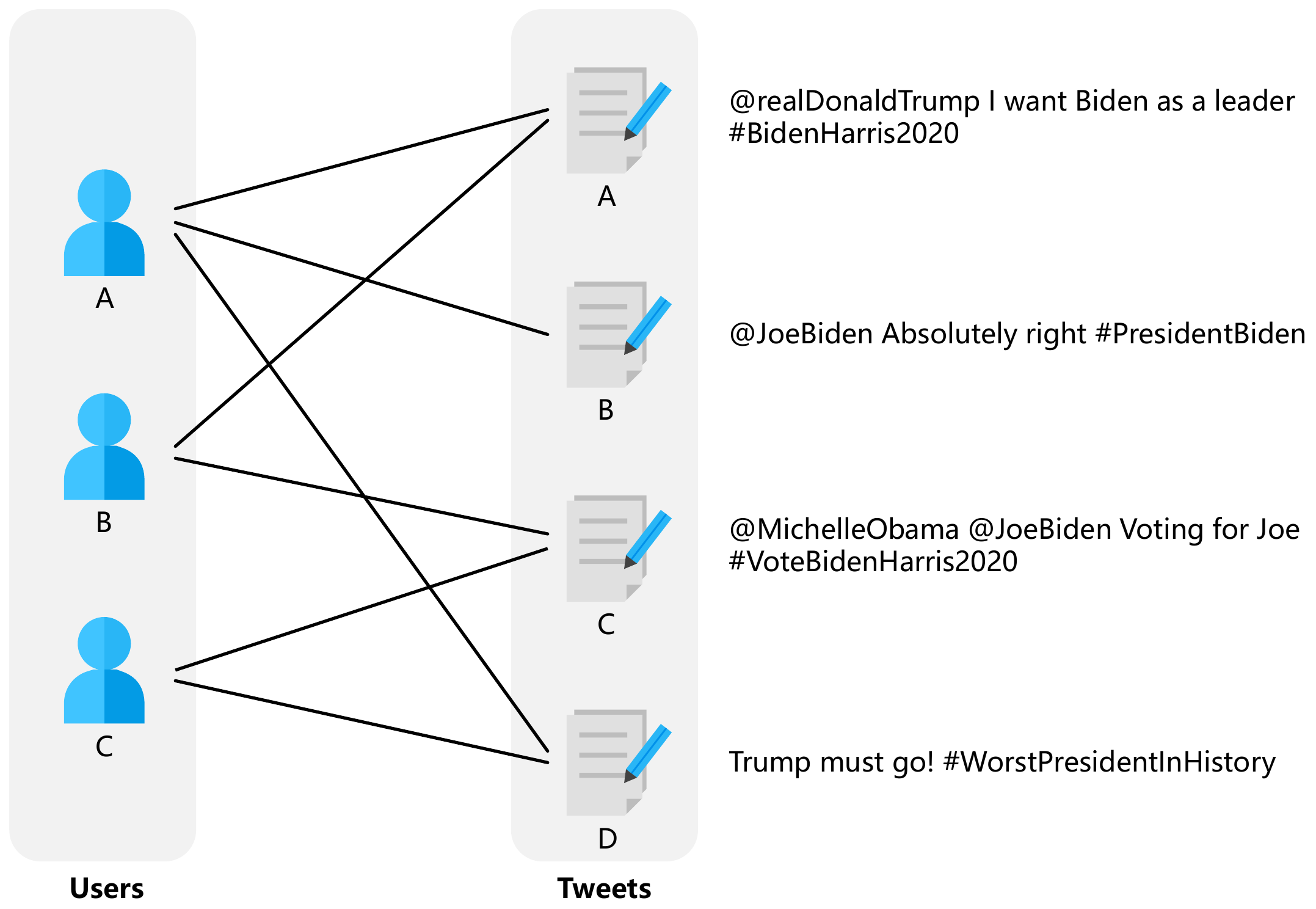}
\caption{ An example of the user-tweet bipartite graph. The links represent posting and retweeting behaviors. }
\label{fig:graph}
\end{figure}

\subsection{User-Tweet Bipartite Graph Construction}
In addition to the content of a tweet, its social context also helps to determine its opinion. Therefore, we build a bipartite graph to model user-tweet interactions, with each edge representing a user's behavior towards a tweet (as illustrated in Figure~\ref{fig:graph}). We consider both posting and retweeting behaviors because users generally retweet messages they ideologically agree with (retweeting is sharing a tweet without adding any comment). The resulting graph contains \num{1123749} tweet nodes, \num{700507} user nodes, \num{1123749} post links, and \num{1473818} retweet links.

Recently, GNNs have been widely used to process such graph-structure data and have shown remarkable performance, so we intend to apply a GNN to learn the representations of tweets from the graph to further determine their opinions. To this end, we first need to assign initial features to the nodes in the graph. Tweet node features can be readily initialized using a pre-trained language model. However, user node features cannot be initialized in the same way due to the lack of textual information. Thus, it is difficult to assign appropriate initial features aligned with tweet nodes to user nodes. 

In fact, in a bipartite graph, each node's first-order neighbors are of a different type than itself, while its second-order neighbors are of the same type as itself. In other words, the second-order neighbors of a tweet node in the user-tweet bipartite graph are also tweet nodes. Furthermore, these tweets tend to share the same opinions as the original tweet. As shown in Figure~\ref{fig:graph}, using users A and B as bridges, tweet A's second-order neighbors are tweets B, C, and D, which express similar opinions as it. To leverage this insight, we introduce a novel skip aggregation mechanism that enables each tweet node to aggregate information from its second-order neighbors. User nodes only serve as bridges during neighborhood sampling, bypassing the initialization problem.

\subsection{Skip Aggregation Graph Convolution Layer} \label{sa}
Most modern GNNs follow a two-phase scheme involving aggregation and combination to update node features in a graph \cite{xu2018powerful}. Formally, the $l$-th layer of a GNN can be defined as:

\begin{equation}
\begin{aligned}
&a_v^{(l)}=\operatorname{AGGREGATE}^{(l)}\left(\left\{h_u^{(l-1)}: u \in \mathcal{N}(v)\right\}\right)
\\
&h_v^{(l)}=\operatorname{COMBINE}^{(l)}\left(h_v^{(l-1)}, a_v^{(l)}\right)
\end{aligned}
\end{equation}

where $a_v^{(l)}$ is the aggregated node feature of node $v$'s neighborhood, $\mathcal{N}(v)$ is a set of neighbor nodes of node $v$, and $h_v^{(l)}$ is the feature of node $v$ at the $l$-th iteration. $h_v^{(0)}=x_v$, where $x_v$ is the initial feature of node $v$.

\begin{figure}[t]
\centering
\includegraphics[width=\linewidth]{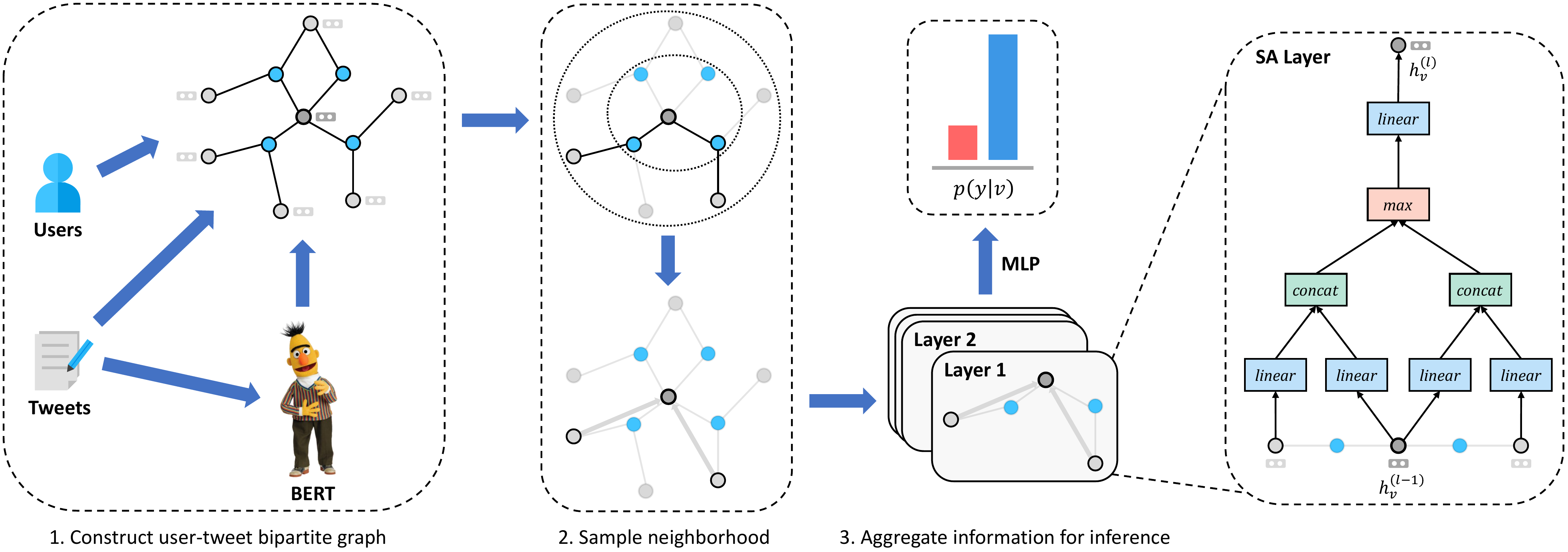}
\caption{ An overview of our proposed framework. A user-tweet bipartite graph is constructed and fed into SA-GNN to infer the opinions of tweets. We illustrate how the SA layer works with an example of a target node aggregating information from two sampled second-order neighbors. The output of the SA layer is a new representation of the target node that combines information from both the target node and its second-order neighbors in the graph. }
\label{fig:model}
\end{figure}

The choice of $\mathcal{N}(\cdot)$, $\operatorname{AGGREGATE}^{(l)}(\cdot)$, and $\operatorname{COMBINE}^{(l)}(\cdot)$ is critical. To achieve efficient training on the large-scale graph, we use a random walk-based strategy for neighborhood sampling. Specifically, for each tweet node in the graph, we simulate $n$ random walks of length 2, recording the visited tweet nodes and their corresponding visit times. The neighborhood of a tweet node is then defined as the top $k$ most visited tweet nodes in the random walks starting at it.

By doing so, we can obtain an informative and representative set of neighbor nodes for each tweet node. The tweet node $v$ and its $k$ neighbor nodes form $k$ unique center-neighbor pairs $\{(v,u):u \in \mathcal{N}(v)\}$. Unlike typical scenarios where only a single edge connects each pair, in this case, each pair is connected by a user node and two edges: one edge $e_{v,u}^v$ links the center node $v$ with the user node, and the other edge $e_{v,u}^u$ links the neighbor node $u$ with the user node. Moreover, due to the heterogeneity of the user-tweet bipartite graph, different center-neighbor pairs can have different types of $e_{v,u}^v$ and $e_{v,u}^u$.

To effectively leverage the information from these two edges, our proposed Skip Aggregation (SA) graph convolution layer takes a different approach than typically aggregating information from neighbor nodes and combining it with the center node's original feature. Instead, the SA layer aggregates features for center-neighbor pairs, considering the types of the two edges between each pair. Formally,
\begin{equation}
\begin{aligned}
&a_v^{(l)}=\operatorname{AGGREGATE}^{(l)} \big(\big\{\sigma\big(\big[\textbf{W}_{cen,\phi(e_{v,u}^v)}^{(l)} \cdot h_v^{(l-1)} \| \textbf{W}_{nei,\phi(e_{v,u}^u)}^{(l)} \cdot h_u^{(l-1)}\big]\big) : u \in \mathcal{N}(v)\big\}\big)
\\
&h_v^{(l)}=\sigma\big(\textbf{W}_c^{(l)} \cdot a_v^{(l)}\big)
\end{aligned}
\end{equation}

\begin{algorithm}[tb]
\caption{SA-GNN embedding generation (i.e., forward propagation) algorithm}
\label{alg:algorithm}
\begin{flushleft}
\textbf{Input}: Bipartite graph $\mathcal{G}(\mathcal{U}, \mathcal{V}, \mathcal{E})$; input features $\left\{\textbf{x}_v: v \in \mathcal{V}\right\}$; depth $L$
\\
\textbf{Output}: Vector representations $\textbf{z}_v$ for all $v \in \mathcal{V}$
\end{flushleft}
\begin{algorithmic}[1] 
\setstretch{1.25}
\STATE $h_v^{(0)} \leftarrow \textbf{x}_v, \forall v \in \mathcal{V}$;
\FOR{$l=1 \ldots L$}
\FOR{$v \in \mathcal{V}$}
\FOR{$u \in \mathcal{N}(v)$}
\STATE $a_{v,u}^{(l)} \leftarrow \sigma\big(\big[\textbf{W}_{cen,\phi(e_{v,u}^v)}^{(l)} \cdot h_v^{(l-1)} \| \textbf{W}_{nei,\phi(e_{v,u}^u)}^{(l)} \cdot h_u^{(l-1)}\big]\big)$;
\ENDFOR
\STATE $a_{v}^{(l)} \leftarrow \operatorname{AGGREGATE}^{(l)} \big(\big\{a_{v,u}^{(l)} : u \in \mathcal{N}(v)\big\}\big)$
\STATE $h_v^{(l)} \leftarrow \sigma\big(\textbf{W}_c^{(l)} \cdot a_v^{(l)}\big)$;
\ENDFOR
\STATE $h_v^{(l)} \leftarrow h_v^{(l)} /\|h_v^{(l)}\|_2, \forall v \in \mathcal{V}$;
\ENDFOR
\STATE $\textbf{z}_v \leftarrow h_v^{(L)}, \forall v \in \mathcal{V}$;
\end{algorithmic}
\end{algorithm}

where $\|$ represents concatenation, $\sigma(\cdot)$ refers to a nonlinear activation function, and $\phi(\cdot)$ is an edge type mapping function. $\textbf{W}_{cen,\phi(e_{v,u}^v)}^{(l)},\textbf{W}_{nei,\phi(e_{v,u}^u)}^{(l)} \in \mathbb{R}^{d \times d}$ are trainable parameters that transform the features of the center node $v$ and the neighbor node $u$ according to the types of $e_{v,u}^v$ and $e_{v,u}^u$, respectively. $\textbf{W}_c^{(l)} \in \mathbb{R}^{2d \times d}$ combines the concatenated features for use in the next layer. The $\operatorname{AGGREGATE}$ operator denotes a versatile aggregator of center-neighbor pair representations that offers several aggregation options, including mean, max, sum, and weighted sum \cite{ying2018graph}.

We refer to the GNN model formed by stacking multiple SA layers as SA-GNN. Algorithm~\ref{alg:algorithm} describes the embedding generation process of $L$-Layer SA-GNN when the bipartite graph $\mathcal{G} = (\mathcal{U}, \mathcal{V}, \mathcal{E})$, where $\mathcal{U}$ and $\mathcal{V}$ are user node and tweet node sets, respectively, and the initial features $\textbf{x}_v$ for tweet nodes are provided as input. The normalization in Line 10 makes training more stable.

\subsection{Model Training}
Figure~\ref{fig:model} shows an overview of our proposed framework. We first construct a user-tweet bipartite graph based on users' posting and retweeting behaviors and initialize the features of tweet nodes using BERT. Then, we sample neighborhood and apply SA-GNN to generate behavior-aware representations for each tweet node. A fully connected layer and a sigmoid function are finally employed to compute the predicted label probabilities.
\begin{equation}
\hat{y}_{v}=\operatorname{Sigmoid}\left(\textbf{W}_o \cdot \textbf{z}_v\right)
\end{equation}
where $\textbf{z}_v$ is the tweet node $v$'s new representation generated by SA-GNN.

We use binary cross-entropy loss to train our model.
\begin{equation}
\mathcal{L}=-\sum_{v \in \mathcal{V}} y_v \log \hat{y_v}+\left(1-y_v\right) \log \left(1-\hat{y_v}\right)
\end{equation}
where $\mathcal{V}$ represents the set of tweet nodes in the graph.

\section{Experiments}
\subsection{Experimental Settings}
\subsubsection{Baselines}
To verify the significance of user behavioral information for detecting political opinions and the effectiveness of skip aggregation in leveraging such information, we compare our model with the following baseline models:
\begin{itemize}
\item
\textbf{BERT} \cite{devlin2018bert} is a powerful language representation model based on the Transformer architecture. It was pre-trained on a large unlabeled corpus using two self-supervised tasks: Masked Language Modeling (MLM) and Next Sentence Prediction (NSP) and can be fine-tuned for various downstream tasks.
\item \textbf{GCN} \cite{kipf2016semi} is a type of neural network designed to perform semi-supervised learning on graph-structured data. It utilizes the symmetrically normalized graph Laplacian to generate node embeddings, which can be viewed as aggregating information from neighbor nodes.
\item \textbf{GAT} \cite{velivckovic2017graph} improves the neighborhood aggregation scheme of GCN by incorporating a self-attention mechanism to assign weights to each node in the neighborhood based on its importance to the target node.
\item \textbf{GraphSAGE} \cite{hamilton2017inductive} is a framework for inductive representation learning on large graphs that can predict the embeddings of new nodes without the need for re-training the entire model.
\item \textbf{GIN} \cite{xu2018powerful} adopts a message passing scheme to update node representations in a way that incorporates information about the entire graph structure, which allows it to capture important structural information that other GNNs may miss.
\item \textbf{R-GCN} \cite{schlichtkrull2018modeling} is an extension of GCN that operates on multi-relational graphs, assigning different transformations to edges of different types and directions.
\end{itemize}
\subsubsection{Implementation Details}
For \textbf{BERT} baseline, we fine-tune the pre-trained BERT\textsubscript{base} model for 5 epochs with a learning rate of 5e-5. The BERT\textsubscript{base} model has 12 layers and 768 hidden units for each layer. 

For graph-based models, we use a pre-trained BERT\textsubscript{base} model to encode the content of tweets as the initial features of the corresponding tweet nodes in the graph. Subsequently, we initialize the user node features for \textbf{GCN}/\textbf{GAT}/\textbf{GraphSAGE}/\textbf{GIN}/\textbf{R-GCN} in three different ways, one random and two heuristic:
\begin{itemize}
    \item \textbf{Random Initialization}: Randomly initialize user node features by sampling from a normal distribution.
    \item \textbf{Centroid-based Initialization}: Inspired by the cluster representation \cite{pal2020pinnersage}, we regard the neighborhood of a user node as a cluster, and take the centroid of the cluster as the user node's initial feature. In other words, each user node's feature is initialized with the average of the features of tweet nodes in its neighborhood.
    \item \textbf{Medoid-based Initialization}: Initialize the user node's feature with the medoid of the cluster, which is the feature of the tweet node with the smallest sum of squared distances in terms of features from other tweet nodes in the neighborhood.
\end{itemize}

Due to the massive graph, it is impossible to fit the features of all nodes into the GPU. Therefore, we perform stochastic mini-batch training by virtue of the neighborhood sampling technique. For graph-based baselines, we randomly select 10 neighbors per layer. For our own model, we sample neighbors using the random walk-based strategy outlined in Section \ref{sa}, simulating 20 random walks and selecting the top 10 most visited tweet nodes per layer. 

\begin{table*}[t]
  \caption{The performance of different models. We conduct five trials with different random seeds and report the mean and standard deviation of the results on the test set. SA-GNN\textsubscript{\textit{w/o edge type information}} is a variant of our model that does not utilize edge type information. Bold denotes the best result, and underline denotes the second-best result.}
  \centering
    \resizebox{\columnwidth}{!}{\begin{tabular}{lccccc}
    \toprule
    Model & \multicolumn{2}{r}{} & Accuracy & F1 score    & AUC \\
    \midrule
    BERT baseline & \multicolumn{2}{c}{BERT} & 90.70 $\pm$ 0.04 & 86.67 $\pm$ 0.07 & 89.43 $\pm$ 0.08 \\
    \midrule
    \multirow{12}[6]{*}{Graph-based baselines} & \multirow{4}[2]{*}{\textit{Random Initialization}} & GCN   & 89.08 $\pm$ 0.05 & 84.26 $\pm$ 0.07 & 87.54 $\pm$ 0.05 \\
          &       & GAT   & 91.41 $\pm$ 0.04 & 87.69 $\pm$ 0.08 & 90.21 $\pm$ 0.10 \\
          &       & GraphSAGE   & 91.42 $\pm$ 0.05 & 87.80 $\pm$ 0.07 & 90.38 $\pm$ 0.05 \\
          &       & GIN   & 90.12 $\pm$ 0.07 & 85.78 $\pm$ 0.11 & 88.70 $\pm$ 0.10 \\
          &       & R-GCN & 91.06 $\pm$ 0.05 & 87.26 $\pm$ 0.08 & 89.95 $\pm$ 0.06 \\
\cmidrule{2-6}          & \multirow{4}[2]{*}{\textit{Centroid-based Initialization}} & GCN   & 91.32 $\pm$ 0.04 & 87.48 $\pm$ 0.09 & 89.95 $\pm$ 0.10 \\
          &       & GAT   & 91.71 $\pm$ 0.03 & 88.09 $\pm$ 0.05 & 90.47 $\pm$ 0.05 \\
          &       & GraphSAGE   & 91.61 $\pm$ 0.06 & 87.94 $\pm$ 0.10 & 90.35 $\pm$ 0.10 \\
          &       & GIN   & 91.40 $\pm$ 0.07 & 87.56 $\pm$ 0.10 & 89.98 $\pm$ 0.09 \\
          &       & R-GCN & 91.63 $\pm$ 0.07 & 87.95 $\pm$ 0.09 & 90.34 $\pm$ 0.07 \\
\cmidrule{2-6}          & \multirow{4}[2]{*}{\textit{Medoid-based Initialization}} & GCN   & 91.36 $\pm$ 0.03 & 87.51 $\pm$ 0.04 & 89.96 $\pm$ 0.04 \\
          &       & GAT   & 91.93 $\pm$ 0.05 & 88.42 $\pm$ 0.08 & 90.73 $\pm$ 0.07 \\
          &       & GraphSAGE   & 91.75 $\pm$ 0.05 & 88.14 $\pm$ 0.07 & 90.49 $\pm$ 0.06 \\
          &       & GIN   & 91.54 $\pm$ 0.07 & 87.78 $\pm$ 0.10 & 90.17 $\pm$ 0.08 \\
          &       & R-GCN & 91.77 $\pm$ 0.03 & 88.16 $\pm$ 0.06 & 90.51 $\pm$ 0.07 \\
    \midrule
    \multirow{2}[2]{*}{Ours} & \multicolumn{2}{c}{SA-GNN} & \textbf{92.73 $\pm$ 0.04} & \textbf{89.62 $\pm$ 0.06} & \textbf{91.73 $\pm$ 0.05} \\
          & \multicolumn{2}{c}{SA-GNN\textsubscript{\textit{w/o edge type information}}} & \underline{92.35 $\pm$ 0.03} & \underline{89.03 $\pm$ 0.05} & \underline{91.20 $\pm$ 0.06} \\
    \bottomrule
    \end{tabular}}%
  \label{tab:table1}%
\end{table*}%

All graph-based models have a hidden size of 768 and are trained for 5 epochs with a learning rate of 1e-3. The AdamW optimizer and a linear learning rate scheduler with 6\% warm-up are adopted to train all models.

We consider Accuracy, F1 score, and AUC as evaluation metrics to quantitatively evaluate model performance. We train the model on the training set and obtain the best model on the validation set, using 80\%, 10\%, and 10\% instances for training, validation, and testing, respectively. To ensure the reliability of our results, we repeat the experiment five times and report the average performance.

\begin{table*}[t]
    \caption{The performance comparison when using different aggregators to aggregate center-neighbor pair representations.}
    \centering
    \begin{tabular}{lccc}
    \toprule
        Model & Accuracy & F1 score & AUC  \\
    \midrule
        SA-GNN\textsubscript{$\operatorname{mean}$} & 92.32 $\pm$ 0.06 & 89.02 $\pm$ 0.08 & 91.25 $\pm$ 0.05 \\
    \midrule
        SA-GNN\textsubscript{$\max$} & \textbf{92.73 $\pm$ 0.04} & \textbf{89.62 $\pm$ 0.06} & \textbf{91.73 $\pm$ 0.05} \\
    \midrule
        SA-GNN\textsubscript{$\operatorname{sum}$} & 92.30 $\pm$ 0.05 & 88.97 $\pm$ 0.07 & 91.18 $\pm$ 0.06 \\
    \midrule
        SA-GNN\textsubscript{$\operatorname{weighted \ sum}$} & 92.16 $\pm$ 0.04 & 88.78 $\pm$ 0.05 & 91.05 $\pm$ 0.03 \\
    \bottomrule
    \end{tabular}
\label{tab:table2}%
\end{table*}

\subsection{Main Results}
As shown in Table~\ref{tab:table1}, our proposed model SA-GNN demonstrates remarkable performance on the test set. Specifically, SA-GNN achieves 92.73\% Accuracy, 89.62\% F1 score, and 91.73\% AUC, consistently outperforming all baseline models. Notably, SA-GNN outperforms BERT, which is fine-tuned for text classification without introducing any additional information, by 2.03\%, 2.95\%, and 2.30\% in terms of Accuracy, F1 score, and AUC, respectively. Moreover, compared with graph-based baselines that incorporate user behavioral information but encounter the user node initialization problem, SA-GNN yields better results across all three random and heuristic initialization strategies. We also consider a variant of our model that does not utilize edge type information, simply aggregating information from second-order neighbors without considering the two edges between each center-neighbor pair, similar to PinSAGE \cite{ying2018graph}. The result shows that the ablated model SA-GNN\textsubscript{\textit{w/o edge type information}} achieves the second-best performance, but still has a significant gap with the complete model in all aspects. Table~\ref{tab:table2} compares the performance of our model when using different aggregators. Among the four aggregators we considered, the $\max$ aggregator proves to be the most effective, further highlighting the importance of selecting the right aggregator.

\subsection{Model Analysis}
\subsubsection{Effect of User Behavioral Information}
To evaluate the significance of user behavioral information for detecting political opinions, we compare graph-based models that incorporate such information with the BERT baseline. The results show that all graph-based models outperform the BERT baseline with the help of user behavioral information, with GCN exhibiting the smallest improvement. This may be due to GCN's inability to selectively aggregate information from neighbor nodes through an attention mechanism or relation type discrimination, which restricts its capacity to represent graph-structured data. In summary, our observations confirm that incorporating user behavioral information is advantageous for political opinion mining.

\subsubsection{Effect of Skip Aggregation}
To investigate the effectiveness of skip aggregation in leveraging user behavioral information, we compare SA-GNN and its variant that both aggregate information from second-order neighbors with graph-based baselines. Among the three initialization strategies, medoid-based initialization is found to be the most effective for graph-based baselines, with GAT achieving the highest performance of 91.93\% Accuracy, 88.42\% F1 score, and 90.73\% AUC. Nevertheless, SA-GNN still demonstrates superior performance over GAT by 0.80\%, 1.20\%, and 1.00\% in terms of Accuracy, F1 score, and AUC, respectively. Furthermore, SA-GNN shows better performance than graph-based baselines even without the use of edge type information. These findings suggest that skip aggregation can not only bypass the user node initialization problem but also make more effective use of user behavioral information.

\subsubsection{Effect of Edge Type Information}
To assess the importance of edge type information and our model's capacity for utilizing such information, we conduct a comparison between SA-GNN and its variant, as well as R-GCN. As presented in  Table~\ref{tab:table1}, there is still a significant gap between SA-GNN and its variant, which confirms our intuition that the information contained in the two edges between the center-neighbor pair is indeed
valuable. Similar trends can also be observed when comparing R-GCN with GCN, which further demonstrates that incorporating edge type information can lead to certain improvements. Moreover, SA-GNN outperforms R-GCN by 0.96, 1.46, and 1.22 points in terms of Accuracy, F1, and AUC, respectively. These results indicate that SA-GNN is more effective in utilizing edge type information and can thus make better use of user behavioral information.

\begin{figure}[t]
\centering
\includegraphics[width=0.9\linewidth]{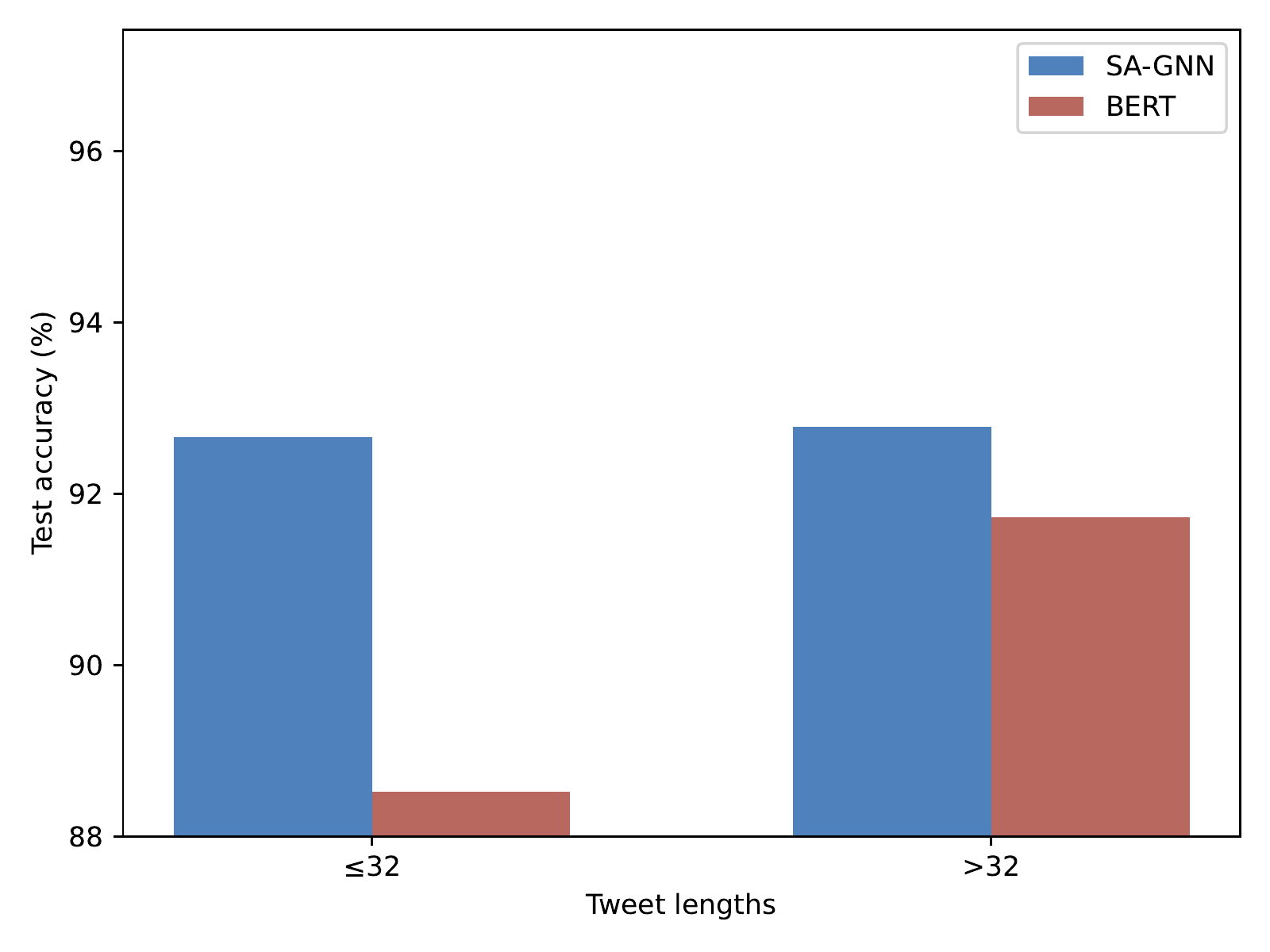}
\caption{The results of performance analysis in terms of tweet length.}
\label{fig:lengths}
\end{figure}

\begin{figure}[t]
\centering
\includegraphics[width=0.9\linewidth]{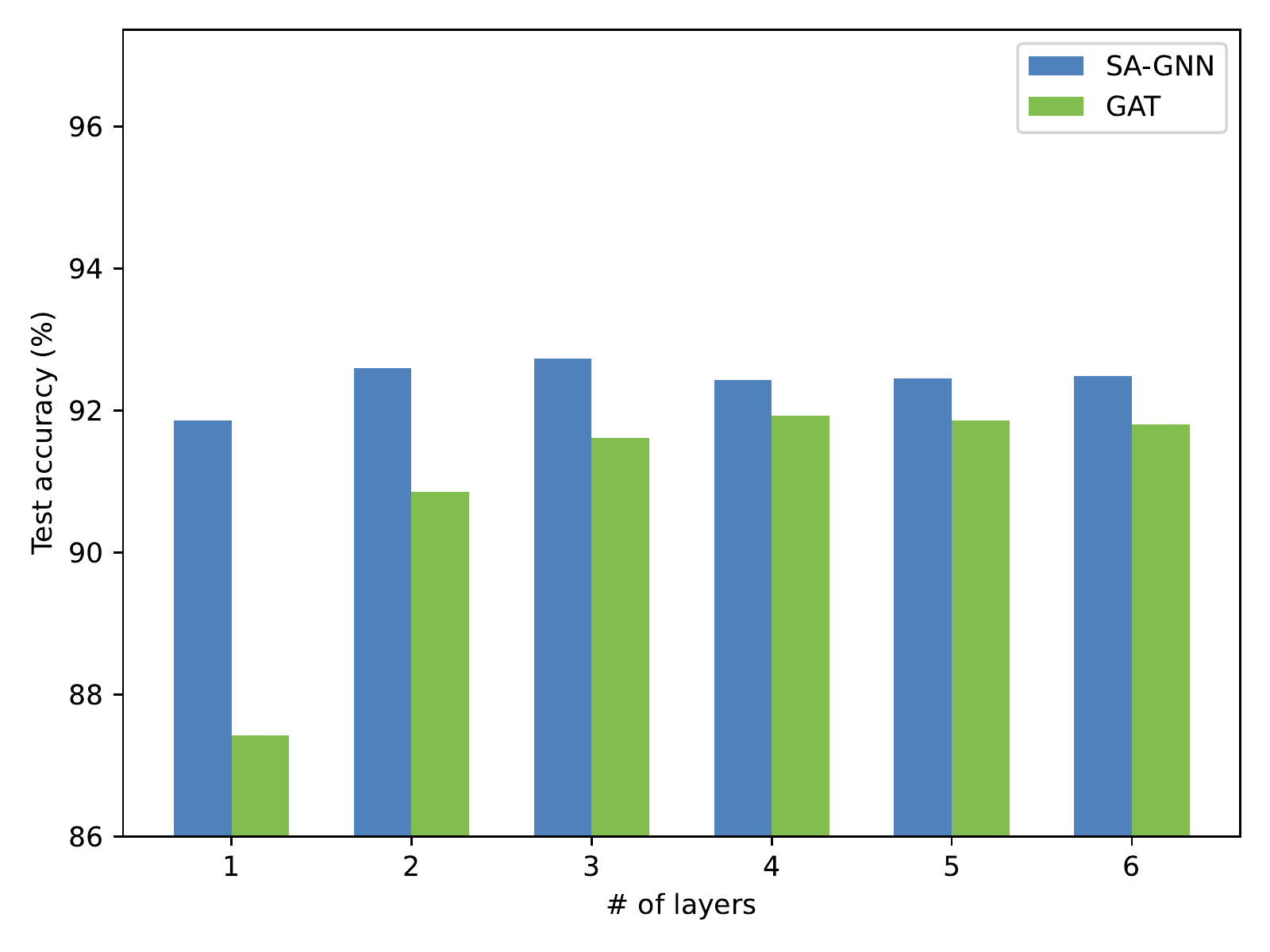}
\caption{The impact of layer number on model performance.}
\label{fig:layers}
\end{figure}

\subsubsection{Performance on Short Texts}
In this part, we aim to conduct a performance analysis to explore our model's performance on short texts. To accomplish this, we first examine the distribution of tweet lengths in the dataset. The length of a tweet is defined as the number of sub-word tokens split by the WordPiece tokenizer used in BERT. Statistics show that 30\% of the tweets in the dataset are no longer than 32 in length, which we consider to be short texts. Then, we compare the performance of SA-GNN and BERT on short texts. As shown in Figure~\ref{fig:lengths}, BERT's performance degrades significantly when dealing with short texts, while SA-GNN is only slightly affected. This can be attributed to the use of user behavioral information in SA-GNN. In general, tweets do not follow strict grammatical rules, and when they are short in length, they contain limited linguistic features and lack context. As a result, it is difficult to determine a tweet's opinion solely based on its content, resulting in BERT's poor performance. However, SA-GNN overcomes this challenge by aggregating information from other relevant tweets through graph convolution, which enriches the representation of short texts and enables the classifier to capture opinions based on sufficient information.

\subsubsection{Impact of SA Layer Number}
We investigate the impact of layer number on the performance of our proposed model SA-GNN and the best-performing graph-based baseline GAT. The layer number is varied from 1 to 6, and the accuracy curves of the two models on the test set are presented in Figure~\ref{fig:layers}. Overall, both models' performance improves as the layer number increases and gradually degrades after reaching a certain depth. The 3-layer SA-GNN achieves the best performance, while GAT performs optimally with 4 layers. Notably, the performance of 1-layer GAT is significantly worse than stacking multiple layers. This observation can be explained by the fact that 1-layer GAT only aggregates information from first-order neighbors, which are all user nodes that contain limited information, resulting in even worse performance than the BERT baseline. Additionally, since SA-GNN aggregates information from second-order neighbors, GAT requires stacking twice as many layers as SA-GNN to capture the same amount of relevant tweet information. However, as the number of layers increases, the vanishing gradient and over-smoothing problems gradually emerge, leading to a drop in performance instead.

\begin{figure}
  \begin{minipage}[t]{\linewidth}
    \centering
    \includegraphics[width=0.8\textwidth]{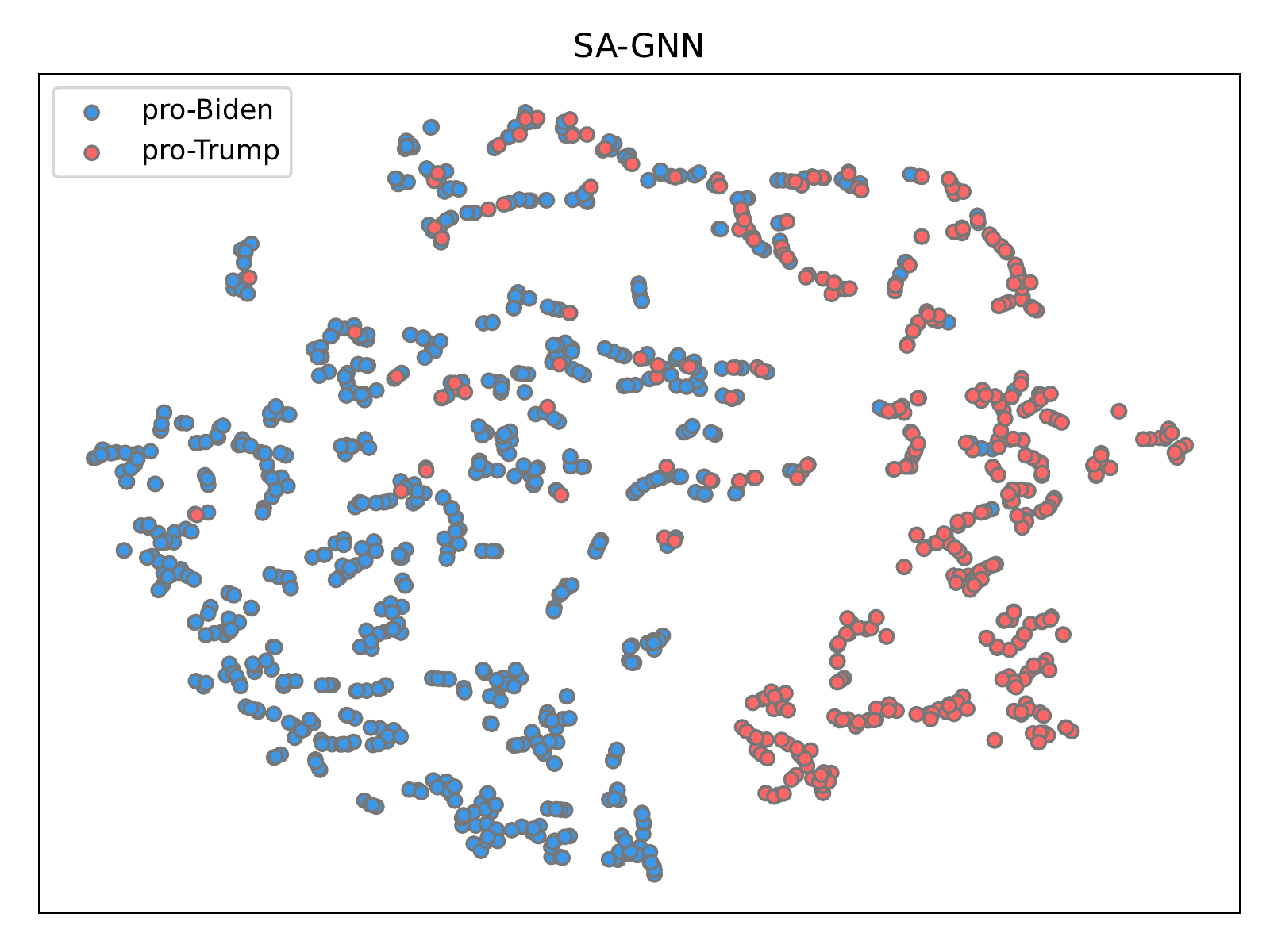}
  \end{minipage}
  \begin{minipage}[t]{\linewidth}
    \centering
    \includegraphics[width=0.8\textwidth]{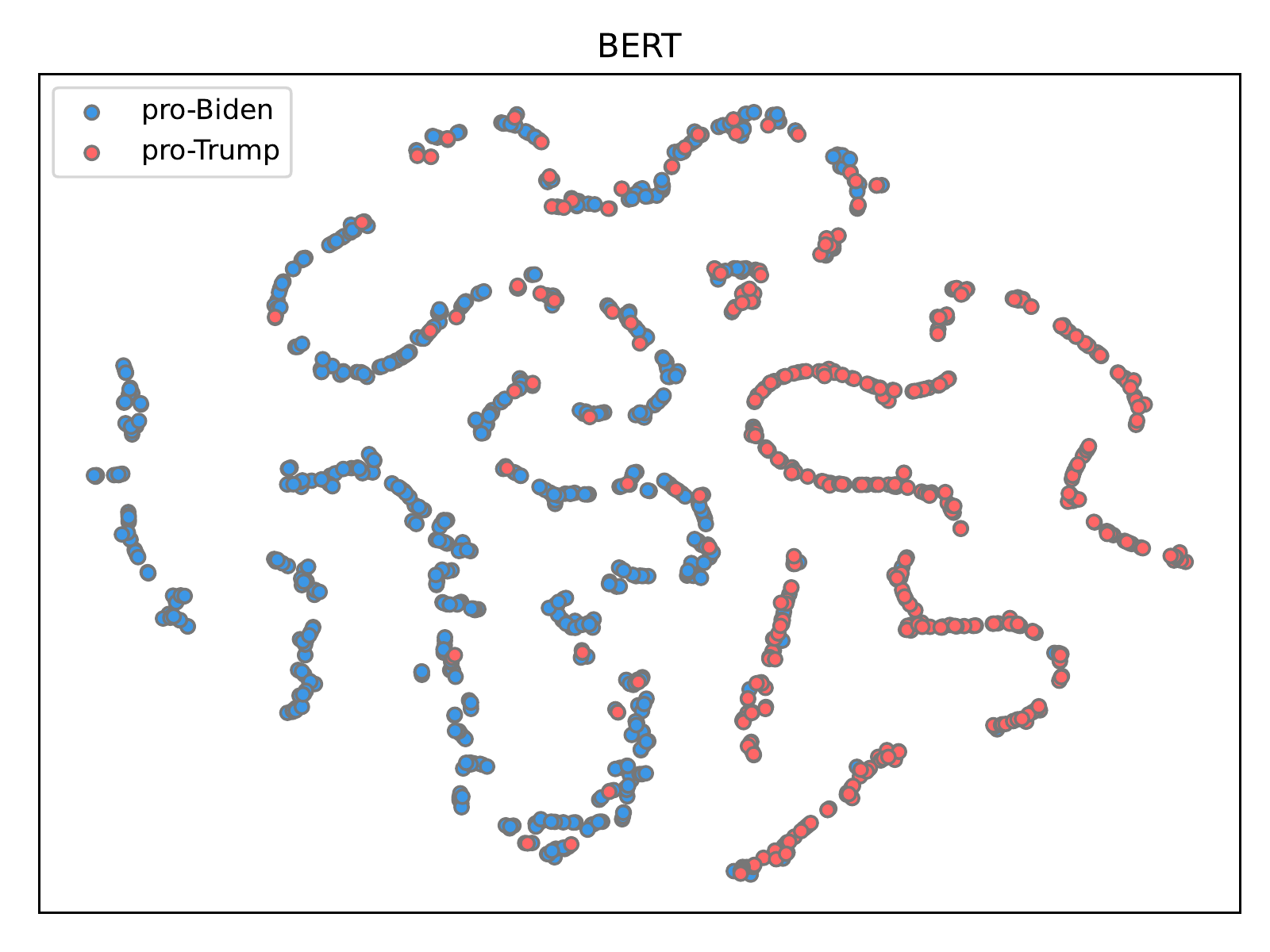}
  \end{minipage}
  \caption{The t-SNE projections of the tweet representations learned by SA-GNN and BERT. The blue and red dots respectively indicate pro-Biden and pro-Trump tweets.}
  \label{fig:visualization}
\end{figure}

\subsection{Embedding Visualization}
To provide a more intuitive verification of our model's effectiveness, we randomly select 1\% of tweets from the test set in a stratified fashion and visualize the t-SNE projections of their representations learned by SA-GNN and BERT in Figure~\ref{fig:visualization}. It can be clearly seen that tweets with the same opinion tend to be more tightly clustered and the boundary between tweets with different opinions is more distinct in the latent space of SA-GNN. This can be attributed to the graph structure used in SA-GNN, which captures the inter-dependencies between different tweets and allows the model to utilize the context and opinions of relevant tweets to generate more distinguishable and higher quality tweet representations than BERT.

\subsection{Misclassification Analysis}
In this section, we investigate and compare the misclassifications made by SA-GNN and BERT on the test set. Specifically, we isolate the samples that are misclassified and analyze the output of the logits layer, which is typically the neural network's final layer for classification tasks and produces raw prediction values.

\begin{figure}[t]
\centering
\includegraphics[width=0.9\linewidth]{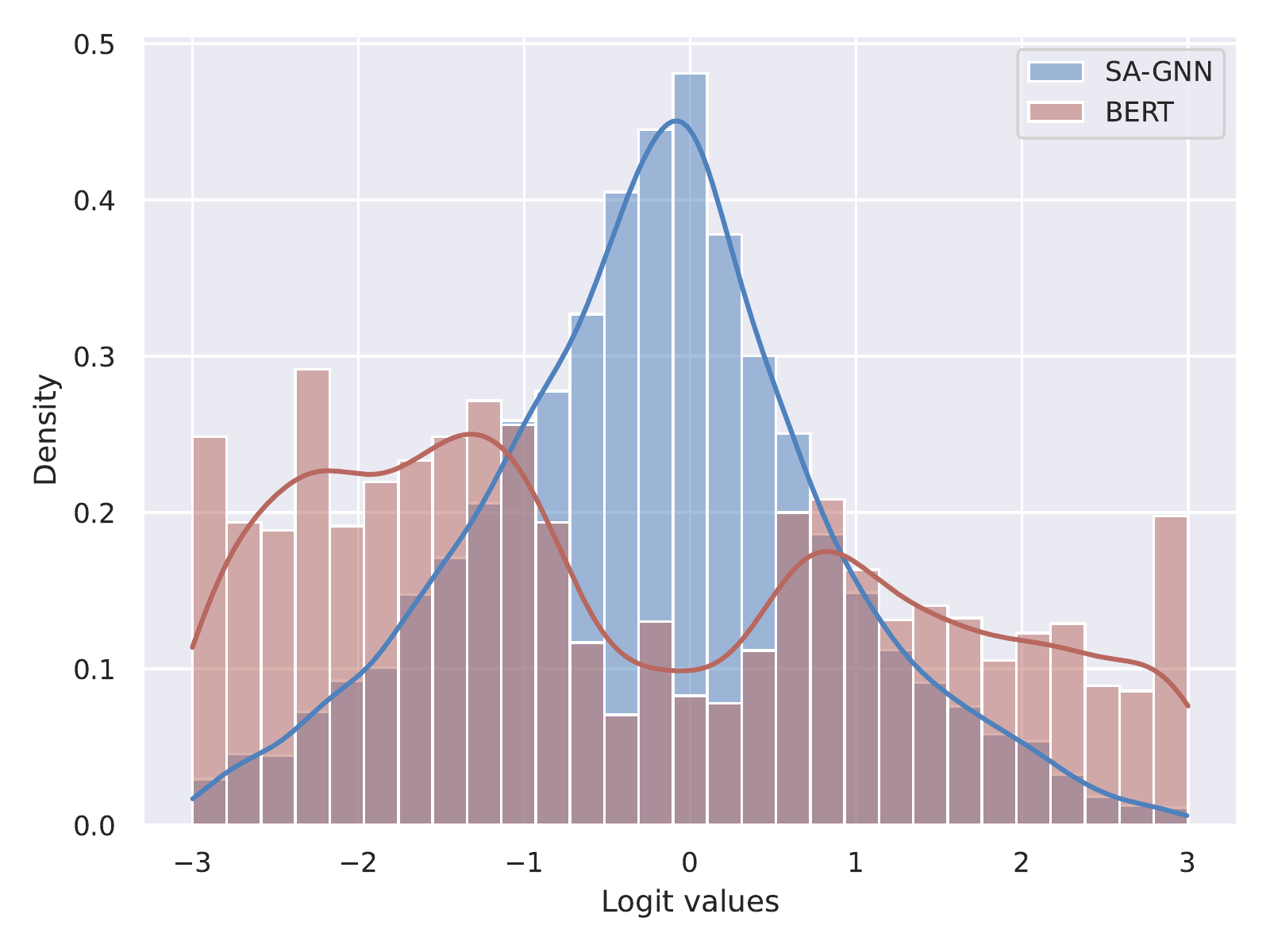}
\caption{ The distributions of logit values output by SA-GNN and BERT for misclassified samples. }
\label{fig:error}
\end{figure}

As illustrated in Figure~\ref{fig:error}, the logit values produced by SA-GNN for misclassified samples are mostly around 0, forming a probability distribution similar to the normal distribution. This may be due to the opinions in most of the tweets misclassified by SA-GNN are implicit or close to neutral, causing the model to make slightly incorrect judgments. In contrast, BERT's logit values for misclassified samples show a bimodal distribution with two peaks at both ends, indicating high confidence on these samples. This could be attributed to BERT determining opinions solely based on content, overfitting to certain common words. Thus, BERT may misinterpret implicit opposing opinions as support or fail to grasp nuances like sarcasm and irony, which are heavily context dependent. Therefore, in addition to superior performance, SA-GNN is less overconfident and more robust than BERT \cite{grabinski2022robust}.

\begin{figure}
\centering
\subfloat[\label{fig:first-order}]{%
\includegraphics[width=0.8\textwidth]{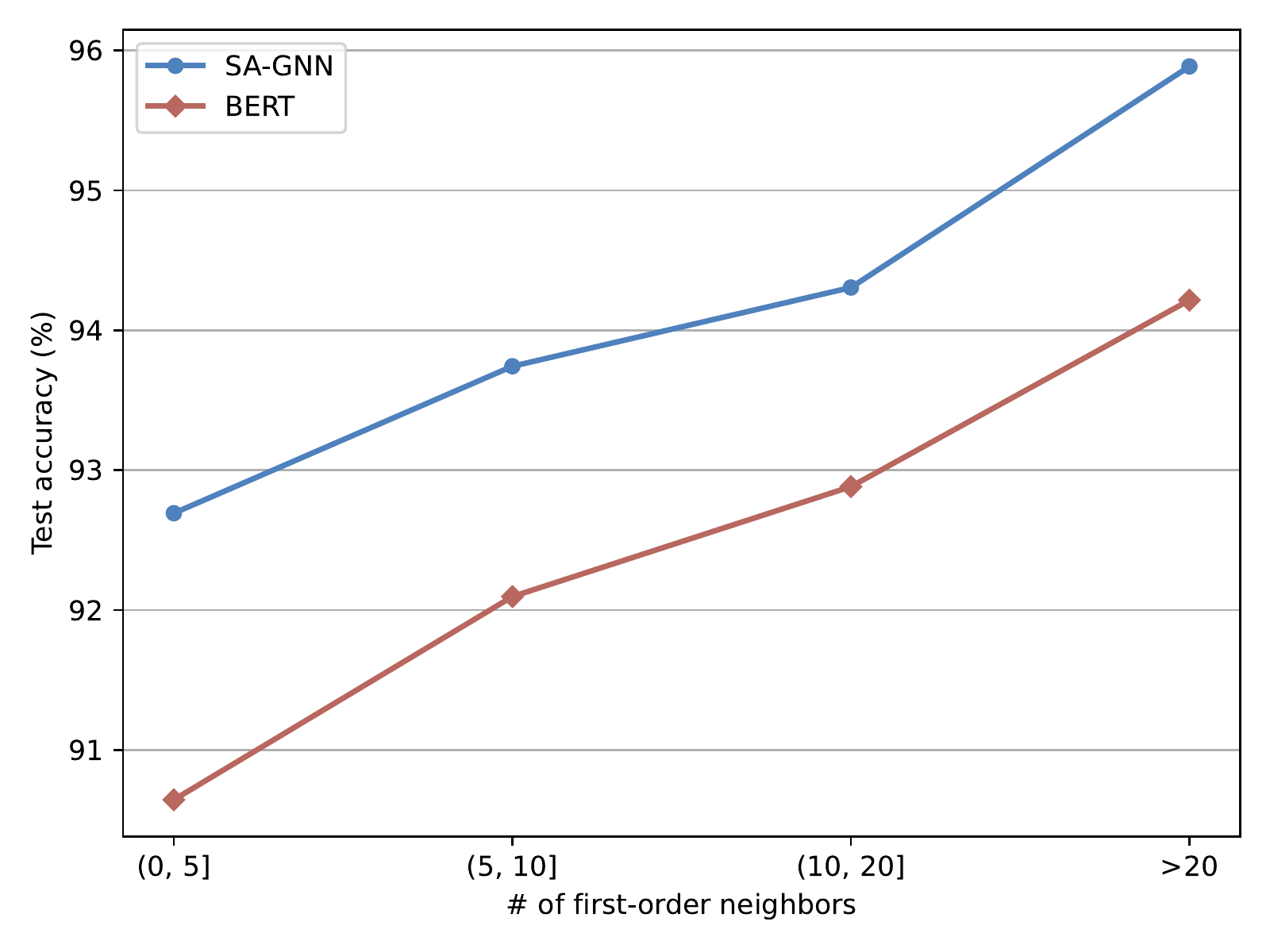}
}
\\
\subfloat[\label{fig:second-order}]{%
\includegraphics[width=0.8\textwidth]{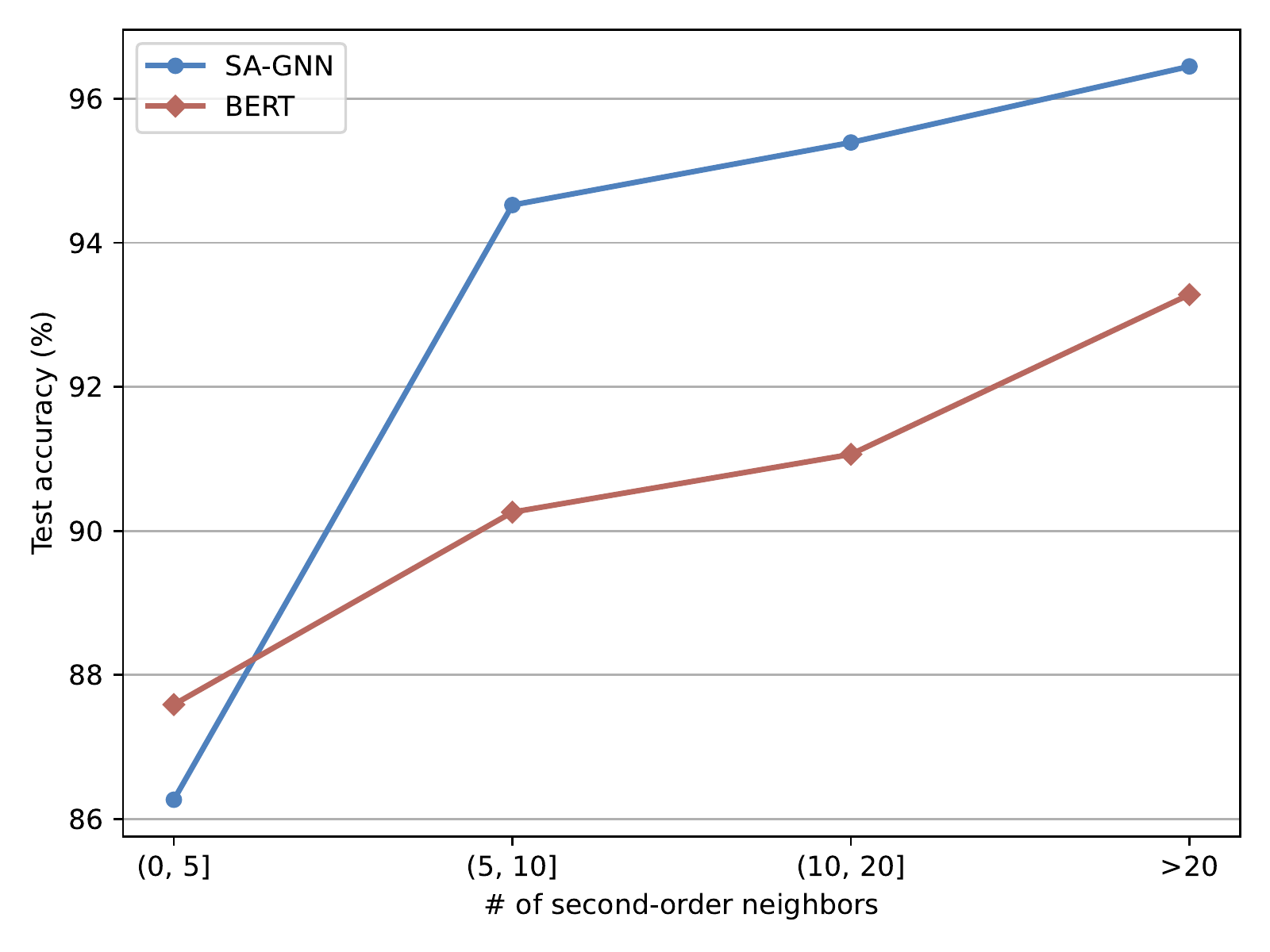}
}
\caption{The results of performance analysis in terms of tweets' (a) first-order neighbor number and (b) second-order neighbor number in the graph.}
\end{figure}

\subsection{Limitations}
Our proposed model employs GNNs to learn behavior-aware tweet representations from the user-tweet bipartite graph, resulting in significant improvements. However, we acknowledge that the model has some limitations. For example, when dealing with tweets that are not popular or whose authors are not active, the model can only aggregate information from a limited number of relevant tweets during the graph convolution phase. As a result, the performance of the model may be degraded. Similarly, for new tweets posted after data collection, their authors or retweeters may not be included in the earlier constructed bipartite graph, in which case the model cannot make inferences properly.

To confirm our intuition, we conduct a performance analysis. Specifically, we compare the performance of SA-GNN and BERT for tweets with different numbers of first-order and second-order neighbors in the graph. As shown in Figures~\ref{fig:first-order} and \ref{fig:second-order}, the performance of BERT, which is unaffected by any graph-related factors, increases with tweets’ first-order neighbor number and second-order neighbor number. This could be because tweets with more neighbors in the graph have more explicit opinions or more formal language, leading the model to be more accurate about them. The performance of SA-GNN also increases with tweets' first-order neighbor number and exhibits the same trend as BERT, indicating that SA-GNN's performance is unaffected by the number of first-order neighbors, as expected because SA-GNN aggregates information from second-order neighbors.

However, the number of second-order neighbors has an impact on SA-GNN's performance. From Figure~\ref{fig:second-order}, we can observe that the performance of SA-GNN degrades dramatically for tweets with no more than 5 second-order neighbors, even far below that of BERT, which is a common problem of GNNs that are biased against low-degree nodes \cite{tang2020investigating}. In fact, although BERT's performance is inferior to graph-based models as it only utilizes the linguistic information of the tweet, it is more flexible and efficient. Hence, we intend to address the limitation in the future by incorporating user behavioral information into BERT, which could be accomplished, for example, by designing new pre-training objectives or contrastive learning, resulting in a more flexible and efficient political opinion mining model while maintaining high performance.

\section{Conclusion}
In this paper, we create a large-scale dataset from Twitter related to the 2020 US presidential election and propose a GNN-based framework that leverages user behavioral information to improve the accuracy of political opinion mining. The framework adopts a novel skip aggregation mechanism to learn behavior-aware tweet representations from a user-tweet bipartite graph, which is shown to be effective in capturing the complex and nuanced nature of political opinions on social media. With further analyses and visualization, we demonstrate the robustness of the framework and showcase the high quality of tweet representations it generates. Finally, we discuss the limitations of this work and suggest new directions for future research.



\bibliographystyle{elsarticle-num} 
\bibliography{main}





\end{document}